\documentclass[twocolumn]{aastex62}
\usepackage{float}
\usepackage{amssymb,amsmath,natbib,gensymb}
\usepackage{color}
\usepackage{ulem}
\usepackage{latexsym}
\usepackage{wasysym}

\shorttitle{$s_{BV}$ -- $s_0^*$ relation}
\shortauthors{Chen et al.}
\begin{document}

\title{A Linear Relation Between the Color Stretch $s_{BV}$ and the Rising Color Slope $s_0^*(B-V)$ of Type Ia Supernovae}
\author[0000-0003-0853-6427]{Ping Chen}\affil{Kavli Institute for Astronomy and Astrophysics, Peking University, Yi He Yuan Road 5, Hai Dian District, Beijing 100871, China.}\affil{Department of Astronomy, School of Physics, Peking University, Yi He Yuan Road 5, Hai Dian District, Beijing 100871, China}\affil{Department of Particle Physics and Astrophysics, Weizmann Institute of Science, Rehovot 76100, Israel}
\author[0000-0002-1027-0990]{Subo Dong}\affil{Kavli Institute for Astronomy and Astrophysics, Peking University, Yi He Yuan Road 5, Hai Dian District, Beijing 100871, China.}

\author{Chris Ashall}\affiliation{Institute for Astronomy, University of Hawaii at Manoa, 2680 Woodlawn Drive, Honolulu, HI 96822,USA}
\author{S.~Benetti}\affil{INAF - Osservatorio Astronomico di Padova, Vicolo dell'Osservatorio 5, I-35122 Padova, Italy}
\author{D.~Bersier}\affil{Astrophysics Research Institute, Liverpool John Moores University, 146 Brownlow Hill, Liverpool L3 5RF, UK}
\author{S.~Bose}\affil{Department of Astronomy, The Ohio State University, 140 West 18th Avenue, Columbus, OH 43210, USA}\affil{Center for Cosmology and Astroparticle Physics, The Ohio State University, 191 W. Woodruff Avenue, Columbus, OH 43210, USA\\}
\author{Joseph Brimacombe}\affiliation{Coral Towers Observatory, Queensland, Australia}
\author{Thomas G.~Brink}\affil{Department of Astronomy, University of California, Berkeley, CA 94720-3411, USA}
\author{David~A.~H.~Buckley}\affil{South African Astronomical Observatory, P.O. Box 9, Observatory 7935, Cape Town, South Africa}
\author{Enrico Cappellaro}\affil{INAF - Osservatorio Astronomico di Padova, Vicolo dell'Osservatorio 5, I-35122 Padova, Italy}
\author{Grant W. Christie}\affil{Auckland Observatory, Auckland, New Zealand}
\author{N.~Elias-Rosa}\affil{INAF Osservatorio Astronomico di Padova, Vicolo dell'Osservatorio 5, 35122 Padova, Italy}\affil{Institute of Space Sciences (ICE, CSIC), Campus UAB, Carrer de Can Magrans s/n, 08193 Barcelona, Spain}
\author[0000-0003-3460-0103]{Alexei V. Filippenko}\affil{Department of Astronomy, University of California, Berkeley, CA 94720-3411, USA}\affil{Miller Senior Fellow, Miller Institute for Basic Research in Science, University of California, Berkeley, CA 94720, USA}

\author[0000-0002-1650-1518]{Mariusz Gromadzki}\affil{Astronomical Observatory, University of Warsaw, Al. Ujazdowskie 4, 00-478 Warszawa, Poland}
\author[0000-0001-9206-3460]{Thomas~W.-S.~Holoien}\altaffiliation{NHFP Einstein Fellow} \affiliation{The Observatories of the Carnegie Institution for Science, 813 Santa Barbara St., Pasadena, CA 91101, USA}
\author{Shaoming Hu}\affil{Shandong Key Laboratory of Optical Astronomy and Solar-Terrestrial Environment, Institute of Space Sciences, School of Space Science and Physics, Shandong University, Weihai, Shandong, 264209, China}
\author{C.~S. Kochanek}
\affil{Department of Astronomy, The Ohio State University, 140 West 18th Avenue, Columbus, OH 43210, USA}
\affil{Center for Cosmology and Astroparticle Physics, The Ohio State University, 191 W. Woodruff Avenue, Columbus, OH 43210, USA\\}
\author{Robert Koff}\affil{Antelope Hills Observatory, 980 Antelope Drive West, Bennett, CO 80102, USA}
\author{Juna A.~Kollmeier}\affil{Observatories of the Carnegie Institution for Science, 813 Santa Barbara Street, Pasadena, CA 91101, USA}
\author{P.~Lundqvist}\affil{The Oskar Klein Centre, Department of Astronomy, Stockholm University, AlbaNova, SE-10691, Stockholm, Sweden}
\author{S.~Mattila}\affil{Tuorla observatory, Department of Physics and Astronomy, University of Turku, FI-20014 Turku, Finland}
\author{Peter A.~Milne}\affil{University of Arizona, Steward Observatory, 933 North Cherry Avenue, Tucson, AZ 85721, USA}
\author{J. A. Mu\~noz}\affil{Departamento de Astronom\'{\i}a y Astrof\'{\i}sica, Universidad de Valencia, E-46100 Burjassot, Valencia, Spain}\affil{Observatorio Astron\'omico, Universidad de Valencia, E-46980 Paterna, Valencia, Spain}
\author{Robert Mutel}\affil{Department of Physics and Astronomy, University of Iowa, Iowa City, IA 52242}
\author{Tim Natusch}\affil{Institute for Radio Astronomy and Space Research (IRASR), AUT University, Auckland, New Zealand}
\author{Joel Nicolas}\affil{Io Variablles-CCD group, 364 chemin de Notre Dame,06220 Vallauris,France}
\author{A.~Pastorello}\affil{INAF - Osservatorio Astronomico di Padova, Vicolo dell'Osservatorio 5, I-35122 Padova, Italy}

\author{Simon Prentice}\affiliation{School of Physics, Trinity College Dublin, The University of Dublin, Dublin 2, Ireland}
\author{J. L. Prieto}
\affil{N\'ucleo de Astronom\'ia de la Facultad de Ingenier\'ia y Ciencias, Universidad Diego Portales, Av. E\'jercito 441, Santiago, Chile}\affil{Millennium Institute of Astrophysics, Santiago, Chile}
\author{Tyler Roth}\affil{Department of Physics and Astronomy, University of Iowa, Iowa City, IA 52242}
\author{B.~J.~Shappee}\affil{Institute for Astronomy, University of Hawaii at Manoa, 2680 Woodlawn Drive, Honolulu, HI 96822,USA}
\author{Geoffrey Stone}\affil{CBA Sierras, 44325 Alder Heights Road, Auberry CA 93602 USA}
\author{K.~Z. Stanek}
\affil{Department of Astronomy, The Ohio State University, 140 West 18th Avenue, Columbus, OH 43210, USA} 
\affil{Center for Cosmology and Astroparticle Physics, The Ohio State University, 191 W. Woodruff Avenue, Columbus, OH 43210, USA\\}
\author{M.~D. Stritzinger}\affil{Department of Physics and Astronomy, Aarhus University, Ny Munkegade 120, DK-8000 Aarhus C, Denmark}
\author[0000-0003-2377-9574]{Todd A. Thompson}
\affil{Department of Astronomy, The Ohio State University, 140 West 18th Avenue, Columbus, OH 43210, USA} 
\affil{Center for Cosmology and Astroparticle Physics, The Ohio State University, 191 W. Woodruff Avenue, Columbus, OH 43210, USA\\}
\author{Lina Tomasella}\affil{INAF - Osservatorio Astronomico di Padova, Vicolo dell'Osservatorio 5, I-35122 Padova, Italy}
\author[0000-0001-6213-8804]{Steven Villanueva}\altaffiliation{Pappalardo Fellow}\affil{Massachusetts Institute of Technology, Cambridge, MA 02139 USA}

\correspondingauthor{Ping Chen}
\email{chenp1220@pku.edu.cn}

\begin{abstract}

Using data from the Complete Nearby ($z_{\rm host}<0.02$) sample of Type Ia Supernovae (CNIa0.02), we discover a linear relation between two parameters derived from the $B-V$ color curves of Type Ia supernovae: the ``color stretch'' $s_{BV}$ and the rising color slope $s_0^*(B-V)$ after the peak, and this relation applies to the full range of $s_{BV}$. The $s_{BV}$ parameter is known to be tightly correlated with the peak luminosity, and especially for ``fast decliners'' (dim Type Ia supernovae), and the luminosity correlation with $s_{BV}$ is markedly better than with the classic light-curve width parameters such as $\Delta{m_{15}(B)}$. Thus our new linear relation can be used to infer peak luminosity from $s_0^*$. Unlike $s_{BV}$ (or $\Delta{m_{15}}$), the measurement of $s_0^*(B-V)$ does not rely on the well-determined time of light-curve peak or color maximum, making it less demanding on the light-curve coverage than past approaches. 

\end{abstract}


\section{INTRODUCTION}

As a supernova (SN) population, SNe Ia show remarkable regularities in observed properties, and in particular, their light curves follow a tight relation between peak luminosity and the decline rate, which is commonly referred to as the width-luminosity relation (WLR, see \citealt{Phillips2017} for a review). The empirical WLR extends from the most luminous 1991T-like SNe Ia \citep{Filippenko1992_1991T, Ruiz-Lapuente1992, Phillips1992} to the least luminous 1991bg-like SNe Ia \citep{Filippenko1992_1991bg, Leibundgut1993AJ_1991bg, Turatto1996}. The WLR not only offers important clues to understanding the physics of the SN Ia population, but also enables using SNe Ia as important cosmological distance indicators. 

\cite{Pskovskii1977, Pskovskii1984} suggested the existence of WLR for SNe Ia by using a post-peak slope parameter, $\beta$, to measure the light-curve decline rate. Its validity  was called into question by some researchers due to concerns over host-galaxy contamination to the SN flux measurements obtained with photographic plates \citep[see, e.g.,][]{Boisseau1991}. \citet{Phillips1993} established the WLR by using well-sampled light curves of nearby SNe Ia observed with charge-coupled devices (CCDs) and introduced the now-classic width parameter $\Delta m_{15}(B)$, which is the $B$-band magnitude difference between the peak $t_{\rm peak}(B)$ and 15 days afterwards. Later, other similar width parameterizations such as the ``stretch'' $s$ were introduced \citep{Perlmutter1997}, and there were also variants like $x_1$ in SALT2 \citep{Guy2007} or $\Delta$ \citep{Jha2007}.

In order to accurately derive $\Delta m_{15}(B)$ directly from the light curve, a good photometric coverage before  $t_{\rm peak}(B)$ to at least 15 days after  $t_{\rm peak}(B)$ is required. Since the light curves evolve slowly over the peak (varying by only $\sim 0.1$\,mag in one week), accurate peak time determination is challenging if the light curve is not densely sampled around the peak. In practice, various template-fitting methods are often employed to derive $\Delta m_{15}(B)$ using well-sampled light-curve templates with known $\Delta m_{15}(B)$ \citep[see, e.g.,][]{Hamuy1995,Hamuy1996b,Prieto2006, Burns2011}. 

For low-luminosity, fast-declining SNe Ia, $\Delta m_{15}(B)$ is found to be a poor width discriminator, and the WLR using $\Delta m_{15}(B)$ shows a large scatter for $\Delta m_{15}(B) \gtrsim 1.7$ mag \citep[see, e.g.,][]{Burns2014, Gall2018}. Similarly, the stretch method fails for fast decliners \citep[see, e.g.,][]{Phillips2017}. From studying high-quality $B-V$ color curves of SNe Ia observed by the Carnegie Supernova Project (CSP), \citet{Burns2014} found that fast-declining SNe Ia reach their reddest $B-V$ color earlier than those with slower decline rates. To better characterize this, \cite{Burns2014} introduced the ``color stretch''  parameter, $s_{BV}$, which is a dimensionless stretch-like parameter defined as $s_{BV} = t_{\rm max}(B-V)/(30~{\rm days})$, where $t_{\rm max}(B-V)$ is time to the maximum (reddest) $B-V$ color with reference to the $B$-band maximum. Since the $B-V$ color quickly declines after reaching maximum,  $t_{\rm max}(B-V)$ corresponds to a sharp ``break'' in the $B-V$ color curve (see Fig~\ref{fig:BVcolor} for examples). Using $s_{BV}$ as a proxy for decline rate (instead of $\Delta m_{15}$), the scatter in WLR significantly reduces at the low-luminosity end, and SNe Ia over the full range of luminosity lie on a tight and continuous correlation \citep{Burns2018}. Moreover, \cite{Ashall2020} found that combining $s_{BV}$ with the time difference between $B$-band and $i$-band maxima is useful to discriminate between SNe Ia subtypes.

Reducing the scatter in WLR using $s_{BV}$ not only has important implications for using SNe Ia as distance indicators, but also sheds new insight into the physics of WLRs. \citet{Wygoda2019} found that $t_{\rm max}(B-V)$ corresponds to an abrupt change in the mean opacities due to ionization-state transitions of $^{56}$Fe and $^{56}$Co in the ejecta, while the timescale to reach the color maximum is determined by the ejecta $^{56}$Ni column density which sets the recombination time of Fe/Co ions. 

To derive $s_{BV}$ from light curves, it is required to measure not only $t_{\rm max}(B-V)$, but also, like $\Delta m_{15}$, a precise estimate of $t_{\rm peak}(B)$ is also needed.

In this work, we report the discovery of a linear relation between $s_{BV}$ and the rising linear slope of $B-V$ color curve of SNe Ia, which is defined as $s_0^*(B-V)$ by us, using the CNIa0.02 sample \citep{Chen2020}. As a proxy for light-curve width and in turn the peak luminosity via WLR, $s_0^*(B-V)$ has the merit of being less demanding on light curve coverage. 

\section{Light-curve Analysis}
\label{sec:measurement}

In this work, we analyze the $B$- and $V$-band light curves presented in \cite{Chen2020}. We include SNe Ia ranging from the most luminous end (i.e, 1991T-like) to the least luminous end (i.e., 1991bg-like) in the analysis, while we exclude several peculiar Ia-like objects that are known to deviate from the WLR (e.g., Iax-type objects such as SN 2017gbb classified by \citealt{Lyman2017} and the 2009dc-like SN Ia-peculiar, ASASSN-15pz studied by \citealt{Chen2019}). In order to obtain accurate $s_{BV}$ measurements, we need to reliably infer the time of the $B$-band peak , $t_{\rm peak}(B)$, and the time of maximum $B-V$ color, $t_{\rm max}(B-V)$. We discuss below how we select the SNe used in the analysis. 

First, the $B$-band light curves need to have coverage around the peak to infer $t_{\rm peak}(B)$. We adopt the criterion of having at least 3 epochs within 6 days of the peak and at least 4 points within 10 days of the peak. $t_{\rm peak}(B)$ is used as the reference time throughout the paper. $t_{\rm peak}(B)$ is obtained with the Gaussian process method if pre-peak data are available, otherwise it is derived from template fitting using SNooPy \citep{Burns2011}. We list the derived $t_{\rm peak}(B)$ and $\Delta m_{15} (B)$ in Table~\ref{tab:lcparam}. 

Then we analyze the $B-V$ color curves for all remaining SNe to infer $s_{BV}$, and the SNe without data late enough to derive $t_{\rm max}(B-V)$ are excluded from the analysis. To obtain reliable $s_{BV}$, we impose the following selection criteria: 1) the $B-V$ color curve needs to have at least 4 data points covering no less than half of the time range between $t_{\rm peak}(B)$ and $t_{\rm max}(B-V)$; 2) after $t_{\rm max}(B-V)$, there need to be at least 2 data points spanning more than 5 days, and the last data point needs to be at least $\sim$\,10 days after $t_{\rm max}(B-V)$. The phase of the $B-V$ color curve is corrected for time dilation. Since all targets have low redshifts, the K-correction is negligible for our analysis and thus is not applied. 

\begin{figure}
\centering
\includegraphics[width=8cm]{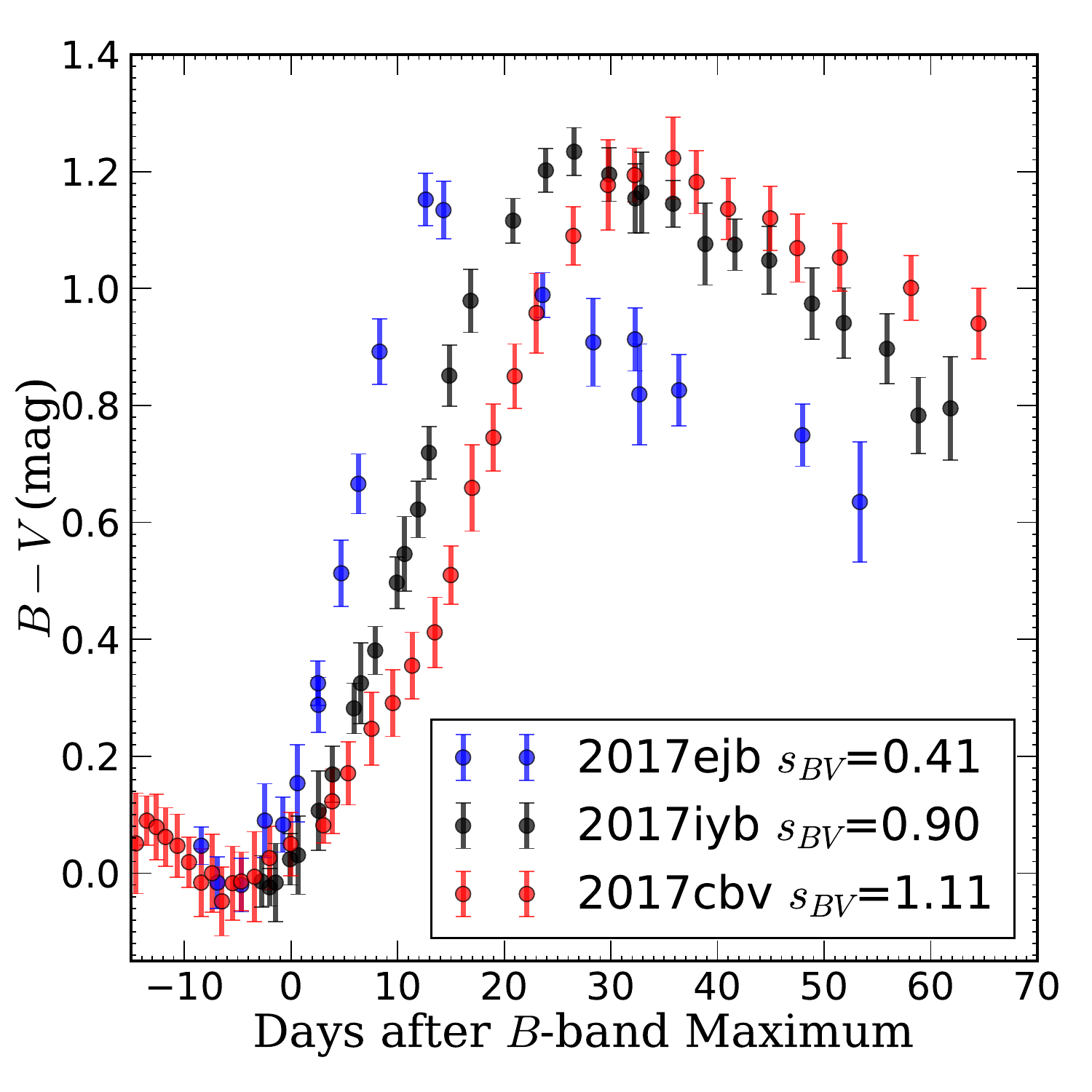}
\caption{$B-V$ color curves of three SNe: 2017ejb, 2017iyb and 2017cbv. The data of 2017ejb is shifted downwards by 0.25 magnitude to match the local minimum of $B-V$ color, $(B-V)_{\rm min}$, of the other two SNe. After $B$-band peak, the $B-V$ color curves have the typical evolution of first rise and then decline. The three SNe reach their maximum $B-V$ colors at different times which correspond to different $s_{BV}$ values (see text for how they are measured). The rising slope seems to be correlated with the time of the maximum $B-V$ color.}
\label{fig:BVcolor}
\end{figure}

In Figure~\ref{fig:BVcolor}, we show the $B-V$ color curves of three SNe with $s_{BV} = 0.41, 0.90, {\rm and}\, 1.11$, respectively. After $t_{\rm peak}(B)$, the $B-V$ color curve soon enters into a linear rise phase (hence, becoming redder) followed by a linear decline (becoming bluer) after $t_{\rm max}(B-V)$. The rising slope and $t_{\rm max}(B-V)$ - $t_{\rm peak}(B)$ seem to correlate with each other, and we measure the relevant light-curve parameters to quantitatively study the observed correlation. In the following, we describe how we measure the parameters from the color curves.

\begin{figure}
\centering
\includegraphics[width=8cm]{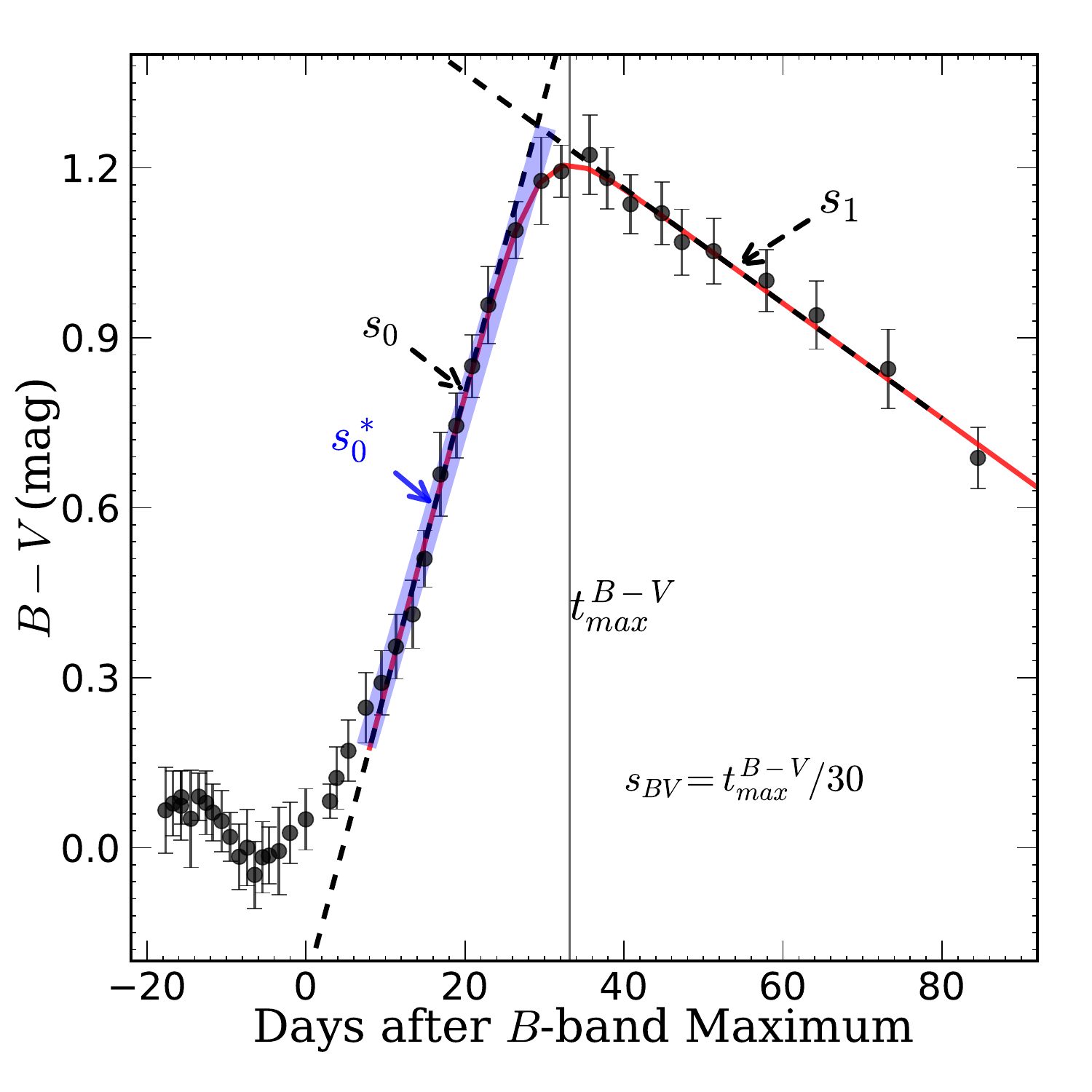}
\caption{$B-V$ color curve of SN 2017cbv with best-fit results shown as an example of the procedure described in this paper. The observed data are shown as black dots and, the best-fit model with  Eq.~(\ref{eq:BVfit_correct}) for the data is shown in red. The black dashed lines indicate the rising slope $s_0$ and declining slope $s_1$ from Eq.~(\ref{eq:BVfit_correct}). The vertical black line marks the time of reddest $B-V$ color. The direct linear fit of the rising slope to get $s_0^*(B-V)$ is shown as the blue line.}
\label{fig:sbv_demo}
\end{figure}

We fit the time evolution of the $B-V$ color since the onset of the post-peak linear rise phase based on Eq.~(2) in \cite{Burns2014}. The model has the form
\begin{equation}
   y(t) = \frac{1}{2}(s_{0}+s_{1})t+\frac{\tau}{2}(s_{1}-s_{0})\ln[\cosh\left(\frac{t- t_{\rm h}}{\tau}\right)]  + c,
\label{eq:BVfit_correct}
\end{equation}
where $s_0$ and $s_1$ characterize the rising and declining slopes, respectively, $\tau$ is the transition time scale indicating how fast the color curve changes from the rising to the declining slope, and $t_{\rm h}$ is the time when the derivative of $y(t)$ equals to the averaged value of the initial and final slopes, i.e, $y'(t_{\rm h})={(s_0+s_1)}{/2}$. We drop the last polynomial term on the right-hand side in Eq.~(2) of \cite{Burns2014}, and that term is used for modeling the color curve prior to the onset of the linear rise phase. In addition, we also correct a typographical error in the first term (correcting ${(s_{0}-s_{1})}/{2}$ in \cite{Burns2014} to $(s_{0}+s_{1})t/2$). Note that, in this model, the rising and declining sides are not strictly linear with time, and $s_0$ and $s_1$ are only supposed to be close to the first derivatives for most parts of the two sides, respectively. We show an example of SN 2017cbv in Figure~\ref{fig:sbv_demo}, in which the best-fit model is shown as a red line. The time of the maximum $B-V$ color where $y'(t_{\rm max}(B-V))=0$, as indicated by the vertical line in Figure~\ref{fig:sbv_demo}, is \footnote{Note that $t_h$ was inaccurately regarded as the time of $B-V$ maximum in \citet{Burns2014} (C. Burns, private communications).} 
\begin{equation}
t_{\rm max}(B-V)=\frac{\tau}{2}\ln\left(-\frac{s_0}{s_1}\right)+t_{\rm h}
\label{eq:truemax}
\end{equation}

We note that the phase when the linear rise starts, $t_s$, is correlated with $t_{\rm max}(B-V)$. We adopt an empirical estimate of $t_s = 0.3 t_{\rm max}(B-V) - 2$ days after $t_{\rm peak}(B)$ as the start phase for our fitting. The adopted end phase for fitting is 90\,days after $t_{\rm peak}(B)$.  Since  $t_{\rm max}(B-V)$ is needed to derive the phase range used for the fit, we iteratively determine this phase range with an initial range of [5, 90]\,days.

We fit the $B-V$ color curves using the Markov chain Monte Carlo (MCMC) method with uniform priors for all parameters. Eq.~(\ref{eq:truemax}) is used to determine $t_{\rm max}(B-V)$ and then $s_{BV}=t_{\rm max}(B-V)/30$. The derived $s_{BV}$, $s_0$ and $s_1$ parameters for 87 SNe Ia with their uncertainties are given in Table~\ref{tab:lcparam}. 

Our work was initially motivated by the apparent correlation between $s_{BV}$ and $s_0$. However, upon further investigations, we find that the value of $s_0$ derived from Eq.~(\ref{eq:BVfit_correct}) depends on the prior placed on the transition timescale $\tau$. The reason for this  dependence can be understood by examining the time derivative of Eq.~(\ref{eq:BVfit_correct}), 
\begin{equation}
   y^{\prime}(t)=\frac{s_{0}+s_{1}}{2}+ \frac{s_{1}-s_{0}}{2}\tanh \left(\frac{t-t_h}{\tau}\right).
   \label{eq:BVslope}
\end{equation}
Clearly the deviation of $y^{\prime}(t)$ from a constant (i.e., as followed from a linear model with respect to $t$) in the rising part of the model depends on $\tau$. In practice, the derived values of $\tau$ often have relatively large uncertainties, and they also correlate with $s_{BV}$, making it difficult to establish a clearcut correlation between $s_0$ and $s_{BV}$.

Instead of using $s_0$ defined in Eq.~(\ref{eq:BVfit_correct}), we measure the rising slope by directly fitting a linear model $y(t) = s_0^*\times t + c^*$ to the rising part of the color curve. To do so, we need to determine a phase range [$t_s$, $t_e$] during which the color curve is well-characterized by a linear model. As can be seen from Figure~\ref{fig:BVcolor}, both the beginning and the end phases of the linear phase seem to be correlated with the time of maximum $B-V$ color. Therefore, the values of $t_s$ and $t_e$ depend on $t_{\rm max}(B-V)$. After some experimentations, we find that the range defined with $t_s = 0.3 t_{\rm max}(B-V) - 2$ and $t_e=0.9 t_{\rm max}(B-V)$ can be well characterized by a linear model for SNe with a broad range of $s_{BV}$. The best-fit linear model within such a range is shown as a blue line in Figure~\ref{fig:sbv_demo}. We exclude SNe with less than 3 points on the $B-V$ color curve within [$t_s$, $t_e$] from the analysis. The directly measured rising color slope parameters, $s_0^*$, are provided in Table~\ref{tab:lcparam}. By comparing the values of $s_0$ and $s_0^*$, we find that $s_0$ tends to be larger than $s_0^*$.

\begin{figure}
\centering
\includegraphics[width=8.5cm]{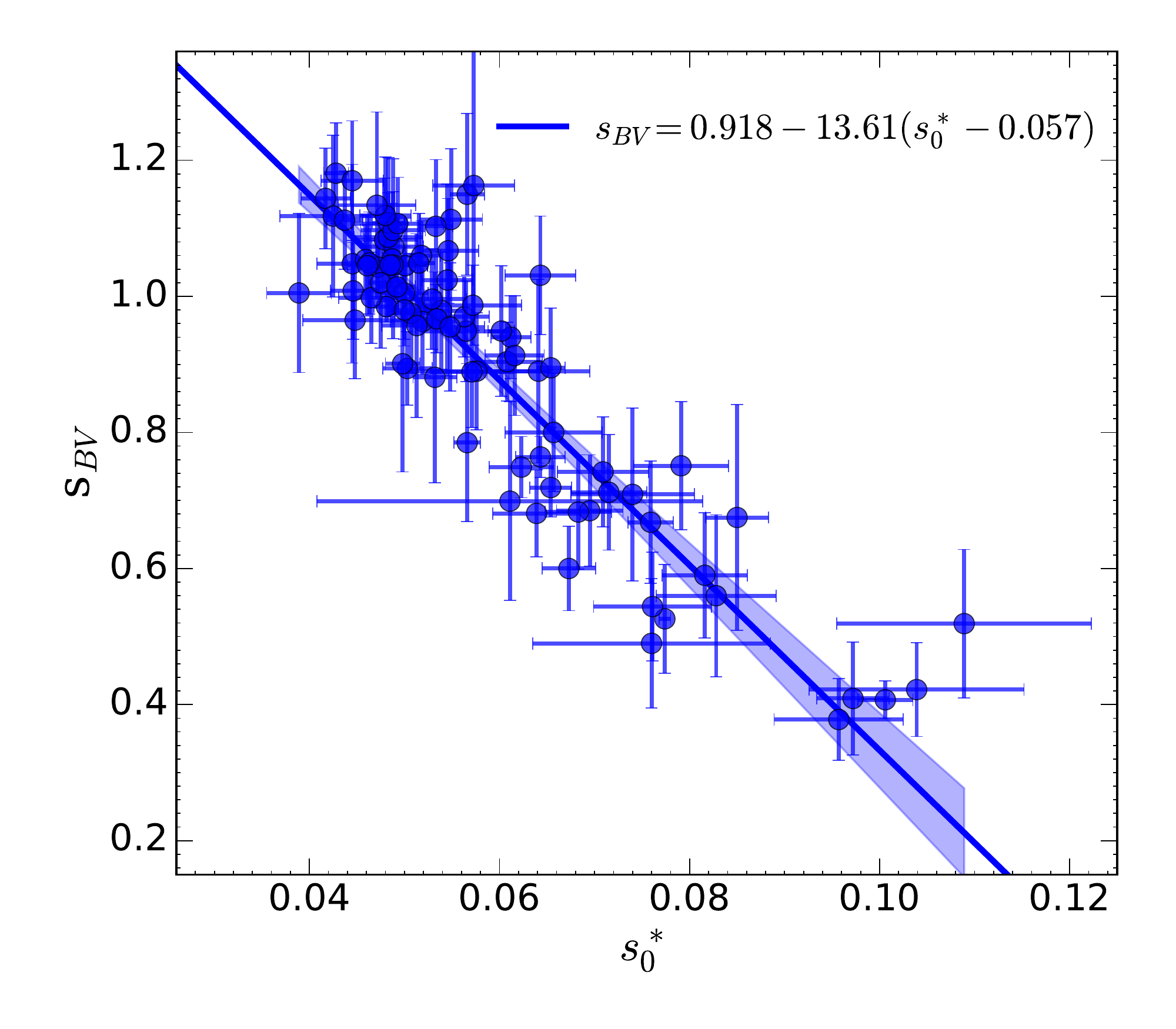}
\caption{Correlation between the color-stretch parameter, s$_{BV}$,  and the rising color slope, $s_0^*$, of the $B-V$ color. All the measurements for 87 SNe in Table~\ref{tab:lcparam} are shown here. The blue line is the best linear fit and the shaded range indicates the 95\% confidence region for the fit.}
\label{fig:sbv_s0s}
\end{figure}

\begin{figure}
\centering
\includegraphics[width=8.5cm]{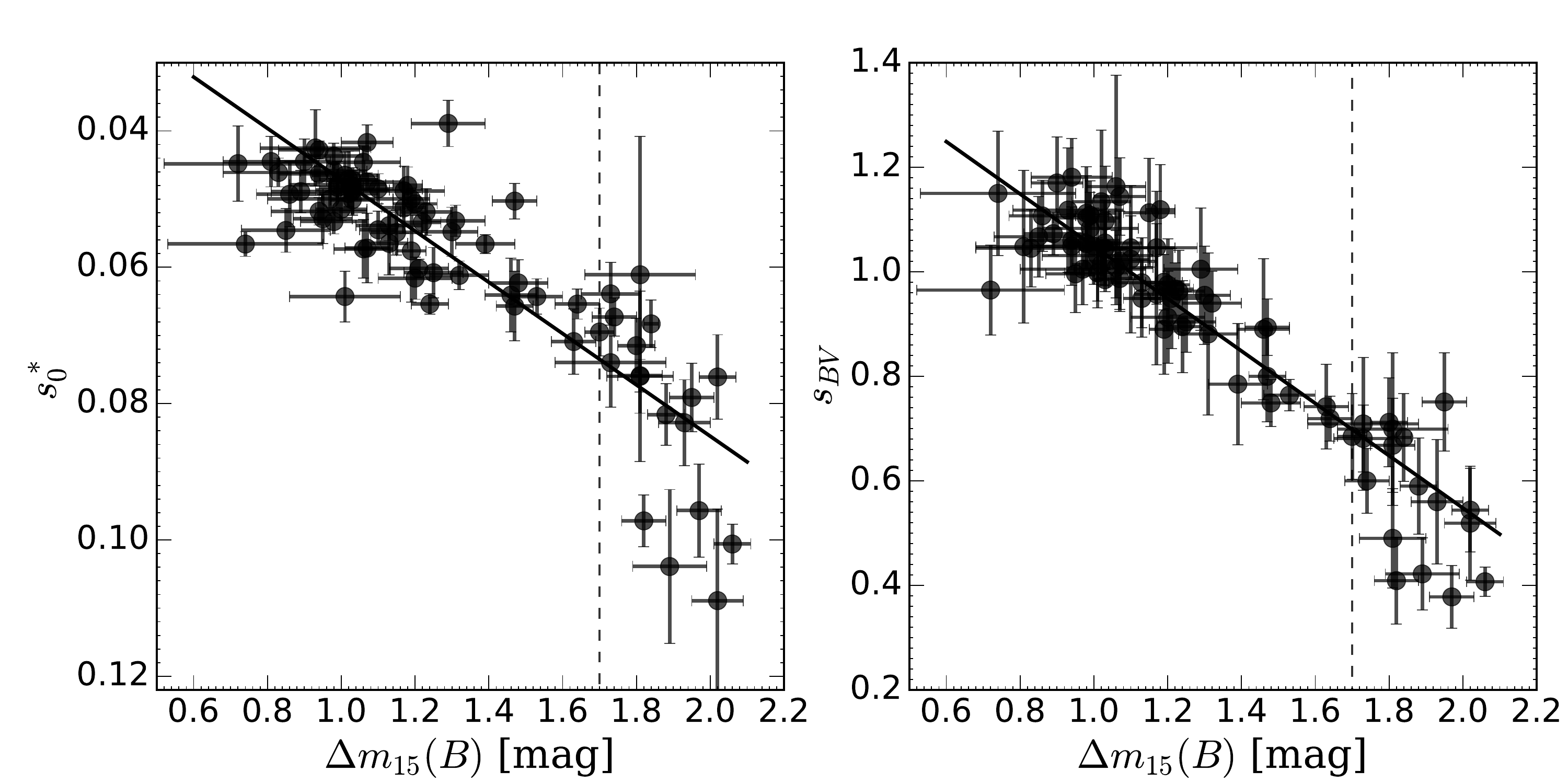}
\caption{Comparisons of  $s_0^*(B-V)$ (left) and $s_{BV}$ (right) with respect to $\Delta{m_{15}(B)}$, respectively. For SNe with $m_{15}(B)<1.7$ (vertical dashed lines), both $s_0^*$ and  $s_{BV}$ are linearly correlated with $\Delta{m_{15}(B)}$ (the black lines in both panels), whereas there is a large scatter for fast decliners with $m_{15}(B)>1.7$.}
\label{fig:m15B_s0s}
\end{figure}

\section{Results} 
\label{sec:result_discussion}
As shown in Figure~\ref{fig:sbv_s0s}, we discover that $s_0^*$ and $s_{BV}$ are linearly correlated, and the linear correlation coefficient is $r=-0.90$, which is highly significant. The best-fit  linear model between $s_0^*$ and $s_{BV}$ accounting for uncertainties yields
\begin{equation}
s_{BV} = 0.918(0.008) -13.61(0.61)\times (s_0^*-0.057).
\label{eq:relation}
\end{equation}
The uncertainties given in the parentheses are obtained by scaling the measurement errors of $s_0^*$ and $s_{BV}$ to have $\chi^2/dof=1$. The residuals to this best-fit linear model have a scatter of 0.093 in $s_{BV}$, which is comparable with the median $s_{BV}$ measurement uncertainty of 0.082. The uncertainties in the $s_0^*$ inference also affect how tight the derived relation is. This linear relation between $s_{BV}$ and $s_0^*$ suggests that $s_0^*$ can be used to infer $s_{BV}$.

As shown in the left panel of Fig.~\ref{fig:m15B_s0s}, we compare $s_0^*(B-V)$ with $\Delta m_{15} (B)$ using our sample. Similar to the well-known trends \citep{Burns2014} when comparing $s_{BV}$ and $\Delta m_{15} (B)$ (right panel of Fig.~\ref{fig:m15B_s0s}), there is a linear correlation for SNe with $\Delta m_{15} (B)<1.7$, but for those with $\Delta m_{15} (B)>1.7$, there is a large scatter.

The inference of $s_0^*$ has the comparative merit over $s_{BV}$ or $\Delta m_{15} (B)$ on the data coverage requirements. Measuring $s_{BV}$ depends on determining two critical timing parameters, $t_{\rm peak}(B)$ and $t_{\rm max}(B-V)$. The measurement of $\Delta m_{15} (B)$ is also highly sensitive to how well $t_{\rm peak}(B)$ is determined. In comparison, deriving $s_0^*$ is simply measuring a slope, which does not require determining either of those two timings, and instead, $s_0^*$ can be determined for a SN which is not observed around the peak and/or lacks the coverage at later times around $t_{\rm max}(B-V)$. 

\section{Measurement Procedure} 
\label{sec:procedure}

In practice, when either $t_{\rm peak}(B)$ or $t_{\rm max}(B-V)$ is absent, it is not possible to use the phase range of  ([$t_s = 0.3 t_{\rm max}(B-V) - 2$, $t_e = 0.9 t_{\rm max}(B-V)$])  with respect to $t_{\rm peak}(B)$ to fit $s_0^*$. Below we provide a practical procedure to measure $s_0^*$ for such light curves:
\begin{itemize}
\item {\it If the $B$-band light curve around the peak is available to measure $t_{\rm peak}(B)$, but there is no sufficiently late observation to determine the break time in the $B-V$ color curve.}  First, start with fitting all data after $t_{\rm peak}(B)$ to get the initial $s_0^*$, and then use Eq.~(\ref{eq:relation}) to estimate $s_{BV}$ and subsequently $t_{\rm max}(B-V) = s_{BV}\times30 \,{\rm d}$. Then use the range of [$0.3 t_{\rm max}(B-V) - 2$, $0.9 t_{\rm max}(B-V)$] for the next fitting and iteratively repeat this process until it converges.  
\item {\it If the $B-V$ color curve around the break time is available to measure the time of maximum $B-V$ color, but there is no $B$-band coverage over the peak to measure $t_{\rm peak}(B)$.} In this case without the reference time of $t_{\rm peak}(B)$, a new reference time of $t_{\rm break}$ (measured as the time of maximum $B-V$ color) will be used. First, start with fitting all data before t$_{\rm break}$ to get the initial $s_0^*$ and then use Eq.~(\ref{eq:relation}) to estimate  $t_{\rm max}(B-V)$. Then use the range from $0.7 t_{\rm max}(B-V) +2$ to $0.1 t_{\rm max}(B-V)$ days before $t_{\rm break}$ for the next fitting and iteratively repeat this process until it converges. 
\item {\it If neither the $B$-band peak nor the break part in the $B-V$ color curve is available.} The $s_0^*$ can be estimated by fitting the available color data that are consistent with a straight line, and this may introduce systematic uncertainties owing to the uncharacterized fitting range. 
\end{itemize}

\section{Summary \& Discussion }
\label{sec:summary}

In summary, we report the discovery of a linear relation between the color stretch parameter, $s_{BV}$, and the rising color slope, $s_0^{*}(B-V)$, of the $B-V$ color after the $B$-band peak, and this relation is applicable to the whole $s_{BV}$ range of SN Ia. With this linear relation, $s_0^{*}$ can be used to infer $s_{BV}$, which is known to be tightly correlated with the peak luminosity, and in comparison, obtaining $s_0^*$ requires a less demanding light-curve coverage than $s_{BV}$.  What also distinguishes $s_0^{*}$ from $\Delta{m}_{15}$ and its variants (e.g., $\Delta{m}_{8}$, $\Delta{m}_{30}$, \citealt{SharonKushnir21}) is that measuring $s_0^{*}$ does not necessarily need coverage over the light-curve peak.  Therefore, applying $s_0^*$ can broaden the capacity of estimating the  luminosity, which is a key parameter of SN Ia.

It will be interesting to study whether $s_0^{*}$ measured in a color other than $B-V$ has similar correlation with $s_{BV}$. In a limited check we perform using the multi-band light curves from the third photometry data release of the first stage of the Carnegie Supernova Project (CSP-I; \citealp{Krisciunas2017}), we find no good correlation between $s_0^{*}(g-r)$ and $s_{BV}$ (or $s_{gr}$). The lack of a good correlation between $s_0^{*}(g-r)$ and $s_{BV}$ (or $s_{gr}$) might be related to the emergence of secondary bumps in the $r$-band light curves of some SNe Ia starting $\sim$ 15 days after $B$-band peak \citep[see, eg.,][]{Papadogiannakis2019}. 

Another future research direction is to study the underlying physical mechanism of the $s_{BV}-s_0^{*}(B-V)$ relation, and it remains to be seen whether it can be interpreted by the same  physical process for WLR via $s_{BV}$ \citep{Wygoda2019} or some other aspect of SN physics.

As shown in \cite{Burns2014, Burns2018}, the WLR via $s_{BV}$ has been successfully used to accurately measure the Hubble constant.  Owing to its correlation with $s_{BV}$,  $s_0^*$ has the potential to be used for cosmology, and since relatively sparse light curves are common for high-$z$ SNe Ia used in cosmology studies, this could be a promising new approach. Further investigations will be needed to establish whether $s_0^*$ will be viable for precision cosmology.

\acknowledgments

We thank Chris Burns for discussing $s_{BV}$. We acknowledge the Telescope Access Program (TAP) funded by the NAOC, CAS and the Special Fund for Astronomy from the Ministry of Finance. We acknowledge SUPA2019A (PI: M.D. Stritzinger) via OPTICON. CSK, KZS and BJS are supported by NSF grants AST-1515927, AST-1814440, and AST-1908570. M.D.S acknowledges funding from the Villum Fonden (project numbers 13261 and 28021). M.D.S is supported by a project grant (8021-00170B) from the Independent Research Fund Denmark. N.E.R. acknowledges partial support from MIUR, PRIN 2017 (grant 20179ZF5KS) and from the Spanish MICINN grant PID2019-108709GB-I00 and FEDER funds. A.V.F.'s supernova group is grateful for financial assistance from the Christopher R. Redlich Fund, the TABASGO Foundation, and the Miller Institute for Basic Research in Science (U.C. Berkeley). A major upgrade of the Kast spectrograph on the Shane 3~m telescope at Lick Observatory was made possible through generous gifts from William and Marina Kast as well as the Heising-Simons Foundation. Research at Lick Observatory is partially supported by a generous gift from Google. We thank the staffs of the various observatories at which data were obtained for their excellent assistance. JLP is provided in part by FONDECYT through the grant 1191038 and by the Ministry of Economy, Development, and Tourism's Millennium Science Initiative through grant IC120009, awarded to The Millennium Institute of Astrophysics, MAS. BJS is also supported by NSF grants AST-1920392 and AST-1911074. MG is supported by the Polish NCN MAESTRO grant 2014/14/A/ST9/00121. Polish participation in SALT is funded by grant no. MNiSW DIR/WK/2016/07. SMH is supported by the Natural Science Foundation of Shandong province (No. JQ201702), and the Young Scholars Program of Shandong University (No. 20820162003).
Support for TW-SH was provided by NASA through the NASA Hubble Fellowship grant \#HST-HF2-51458.001-A awarded by the Space Telescope Science Institute, which is operated by the Association of Universities for Research in Astronomy, Inc., for NASA, under contract NAS5-26555.

\startlongtable
\begin{deluxetable*}{lcccccc}
\tabletypesize{\footnotesize}
\tablecolumns{7}
\tablewidth{0pt}
\tablecaption{Light Curve Parameters of SNe Ia in CNIa0.02 sample\label{tab:lcparam}}
\tablehead{
\colhead{SN} & 
\colhead{t$_{\rm peak}$($B$)}   & 
\colhead{$\Delta m_{15}(B)$} & 
\colhead{s$_0$} &
\colhead{s$_1$}  &
\colhead{$s_{BV}$} &
\colhead{s$_0^*$}  \\
\colhead{} &
\colhead{(MJD)} & 
\colhead{(mag)} & 
\colhead{} &
\colhead{} &
\colhead{} &
\colhead{}
}
\startdata
2016bfu & 57469.7$\pm$0.6 & 1.81$\pm$0.09 & 0.109$\pm$0.019 & $-0.0129\pm$0.0053 & 0.49$\pm$0.09 & 0.076$\pm$0.012 \\
2016blc & 57489.8$\pm$1.1 & 0.99$\pm$0.13 & 0.050$\pm$0.005 & $-0.0099\pm$0.0014 & 1.08$\pm$0.09 & 0.048$\pm$0.003 \\
2016fff & 57630.0$\pm$0.2 & 1.81$\pm$0.06 & 0.082$\pm$0.009 & $-0.0119\pm$0.0039 & 0.67$\pm$0.09 & 0.076$\pm$0.002 \\
2016fej & 57635.5$\pm$0.7 & 0.89$\pm$0.08 & 0.049$\pm$0.002 & $-0.0125\pm$0.0007 & 1.07$\pm$0.04 & 0.049$\pm$0.003 \\
2016gtr & 57667.1$\pm$0.9 & 0.94$\pm$0.11 & 0.045$\pm$0.005 & $-0.0204\pm$0.0068 & 1.18$\pm$0.07 & 0.043$\pm$0.001 \\
\enddata
\vspace{0.1cm}
\textbf{Notes.} This table is available in its entirety in machine-readable format in the online journal. A portion is shown here for guidance regarding its form and content.
\end{deluxetable*}

\newpage
\bibliographystyle{apj}
\bibliography{ms}

\end{document}